\documentclass[
  reprint,
  superscriptaddress,
  amsmath,amssymb,
  aps,
  prmaterials,
]{revtex4-2}

\usepackage[T1]{fontenc}
\usepackage[utf8]{inputenc}

\usepackage{xcolor}
\usepackage{graphicx}
\usepackage{epstopdf}
\usepackage{dcolumn}
\usepackage{bm}
\usepackage{amsfonts}
\usepackage{upgreek}
\usepackage{multirow}
\usepackage{tabularx}
\usepackage{wrapfig}
\usepackage{booktabs}
\usepackage[section]{placeins}
\usepackage[colorlinks=true,linkcolor=blue,citecolor=blue,urlcolor=blue]{hyperref}

\begin{document}

\author{Frank M. Abel}
\email{abel@usna.edu}
\affiliation{National Institute of Standards and Technology, Gaithersburg, MD 20899, USA}
\affiliation{Physics Department, United States Naval Academy, Annapolis, MD 21402, USA}

\author{Paige Burke}
\affiliation{Physics Department, United States Naval Academy, Annapolis, MD 21402, USA}

\author{Daniel Wines}
\affiliation{National Institute of Standards and Technology, Gaithersburg, MD 20899, USA}

\author{Brian Donovan}
\affiliation{Physics Department, United States Naval Academy, Annapolis, MD 21402, USA}

\author{Michelle E. Jamer}
\affiliation{Physics Department, United States Naval Academy, Annapolis, MD 21402, USA}

\author{Kamal Choudhary}
\affiliation{Johns Hopkins University, Baltimore, MD, 21218, USA}
\affiliation{National Institute of Standards and Technology, Gaithersburg, MD 20899, USA}

\date{\today}

\begin{abstract}
Automation and high-throughput characterization and synthesis for material development are becoming increasingly common; these approaches require machine learning (ML) tools to assess material properties, ideally based on a single measurement. Here, ML models are developed to predict magnetization from X-ray diffraction (XRD) for iron oxide nanoparticles. Our approach is to first develop a set of simulated data that links modulated XRD, based on a crystallographic information file (CIF), to a simple magnetic model to determine magnetization at a given magnetic field, thereby enabling us to train Random Forest and Gradient Boosting regression models on a large amount of simulated data. The models are validated by synthesizing iron oxide nanoparticles and measuring their crystal structure via XRD and room-temperature magnetization curves. In doing so, we can fine-tune both the training hyperparameters and the optimal size of the simulated datasets used to train the models. Through this optimization, the best models can achieve an $R^2$ greater than 0.9 for five experimental samples, used for tuning, for predicting the max magnetization (at 2.8 T) of the measurement. Lastly, we demonstrate reasonable predictions on the full magnetization vs. magnetic field curve, showing that the RF model excels at predicting the high magnetic field values, which is key for determining the success of an iron oxide nanoparticle synthesis for applications like magnetic particle imaging (MPI), thermal magnetic particle imaging (T-MPI), and hyperthermia.  
\end{abstract}

\title{Machine Learning for Predicting Magnetization from X-ray Diffraction of Iron Oxide Nanoparticles Using Simple Physics-Based Data Generation}

\maketitle

\section*{Introduction}

The push for increased levels of autonomous material characterization, synthesis, and discovery has increased the need for integration of machine learning (ML) models along with high-throughput experimentation. Self-driving laboratories (SDLs) aim to accelerate the fundamental material science loop of synthesis, characterization, and process optimization by automating certain or all aspects of experimental workflow. \cite{macleod2020self, vriza2023self, lunt2024modular} X-ray diffraction (XRD) is a non-destructive technique that is one of the core and often times first characterization methods used to evaluate the success of an experiment aimed to synthesize a certain material structure. XRD can determine a material's crystalline phases, crystallite size, and structural ordering, which often strongly correlates with the functional properties. In the context of autonomous experimentation, XRD has been shown to serve as a feedback mechanism for assessing the structure of synthesized materials. \cite{szymanski2023autonomous} Other work has developed ML-based approaches to analyze and identify phases in XRD patterns. \cite{szymanski2024integrated, fei2025dara} Additionally, the general inverse problem of solving crystal structures from powder XRD has been shown to be extremely challenging even when constrained to a specific crystal family.\cite{segal2025losslandscapepowderxray} In magnetic materials, especially those at the nanoscale, crystallographic features encoded in XRD patterns are often linked to functional magnetic properties, such as saturation magnetization (M$_s$) and coercivity (H$_c$). Traditionally, these properties are measured using magnetometry, but these measurements can be slow and less amenable to real-time feedback, especially with the measurement of a high volume of samples. Leveraging ML to extrapolate magnetic information from XRD patterns could dramatically increase experimental throughput and decision-making, particularly in the optimization of magnetic material synthesis.
\par
Recent studies have explored ML methods to predict magnetic properties in a number of limited circumstances. In the case of Ce-doped Nd$_2$Fe$_{14}$B magnets, a ML model was trained on a combination of density functional theory (DFT), experimental values from literature, and micromagnetic derived descriptors to make realistic predictions of coercivity. \cite{bhandari2024accurate} Qiao et al. \cite{qiao2023performance} applied interpretable ML to relate composition and sintering process parameters to the performance (coercivity and remanence) of NdFeB magnets. Similarly, Kovacs et al. \cite{kovacs2023physics} combined experimental data and micromagnetic simulations in a physics-informed ML model to guide the design of Nd-lean magnets. Graph Neural Networks (GNNs) trained on magnetic simulation data have been developed to predict the coercivity of hard magnetic microstructures \cite{moustafa2025graph}, and ML has been used to analyze the microstructure of permanent magnets to pinpoint the characteristics that cause magnetization reversal using simulated data and micromagnetic validation. \cite{exl2018magnetic} Other efforts, such as Liu et al. \cite{liu2022magnetic}, employed ML to link composition and process parameters to magnetic properties in substituted Sr-hexaferrites. ML has also been used to predict magnetic anisotropy and volume fraction of defects in FePt thin films. \cite{dengina2022machine} Lastly, ML methods have been used to accelerate the discovery of new magnetic materials by using DFT for structures, energies, and magnetic properties, allowing the prediction of promising material candidates that can be validated experimentally. \cite{xia2022accelerating}
\par
\begin{figure}
    \centering
    \includegraphics[width=\columnwidth]{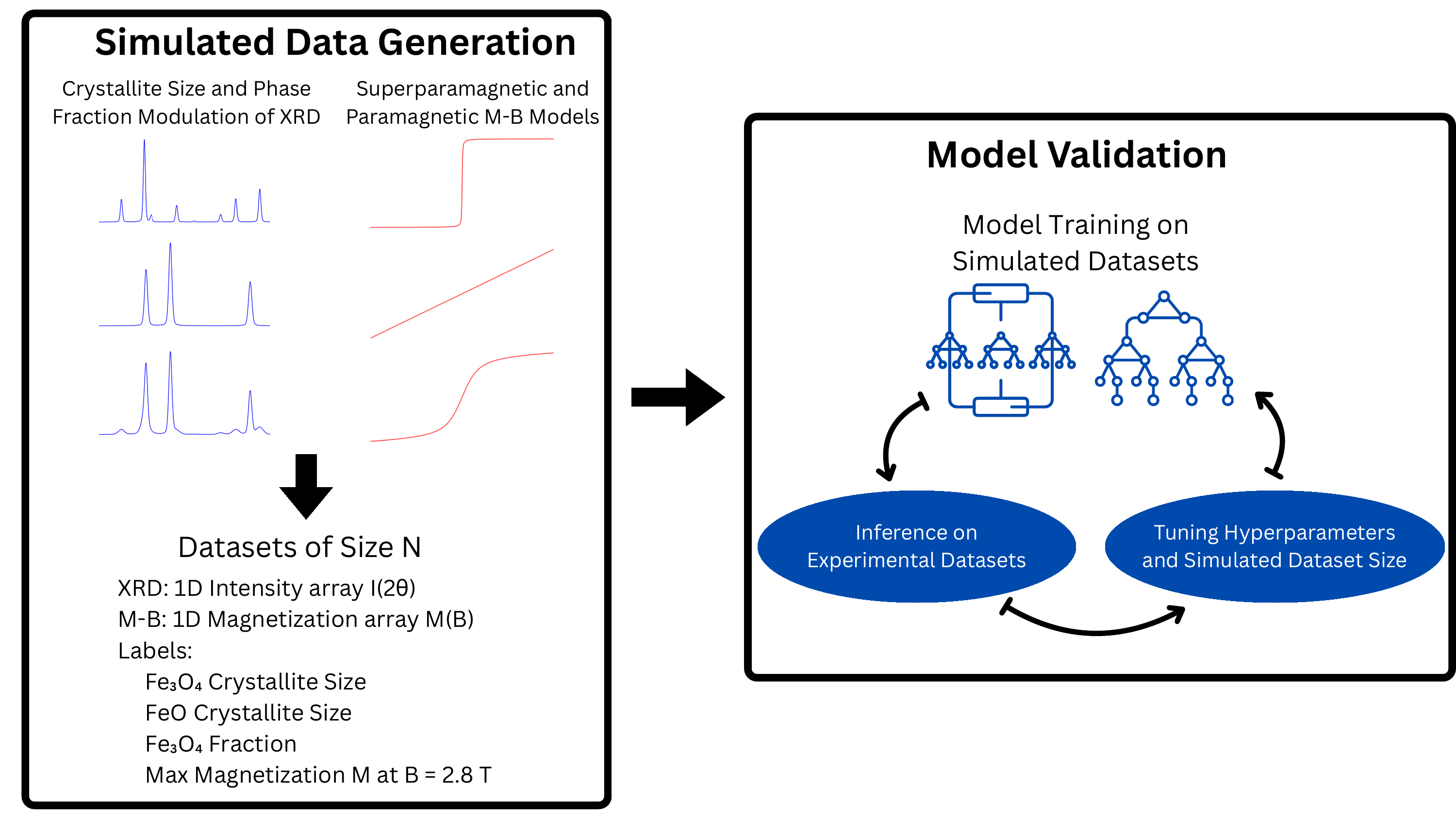}
    \caption{Workflow schematic of simulated data generation, model validation, and tuning based on inference using experimental datasets.}
    \label{fig:Fig1}
\end{figure}
Despite these advances, there are no current studies that use simulated XRD patterns as the sole input for ML-based prediction of magnetization and that are validated by synthesis and characterization of materials relevant to the model. Most methods rely on custom features, micromagnetic simulations, DFT, direct experimental measurement, or a combination of the above. In this work, we propose a generalizable strategy that links structure and property by training models directly on simulated XRD data paired with simulated magnetization vs. magnetic field data with a set of shared parameters. Our method generates simulated datasets from modulated XRD using peak positions and intensities of each phase from crystallographic information files (CIFs), allowing a wide range of configurations, such as phase mixing and different crystallite sizes, to be considered. This enables correlation with simple magnetic models to determine the magnetization as a function of magnetic field for a set of shared parameters. We apply this approach to iron oxide nanoparticles (Fe$_3$O$_4$ and FeO), where for a given XRD pattern, the magnetization vs. magnetic field can be generated using a Langevin function for the Fe$_3$O$_4$ and a linear function for the FeO mimicking its paramagnetic nature at room temperature. Iron oxide has been chosen as a testbed material for our approach due to its simplicity to model and its critical role in technologies, such as magnetic particle imaging (MPI), thermal magnetic particle imaging (T-MPI) \cite{abel2023thermosensitivity, bui2023harmonic}, and magnetic hyperthermia \cite{castellanos2021milestone}. Additionally, in recent years it has been demonstrated that high crystal quality iron oxides (nearly single phase Fe$_3$O$_4$) of a certain particle size can lead to significant enhancement in signal and imaging resolution for MPI \cite{tay2021superferromagnetic, fung2023first, abel2024strongly, bui2024magnetodynamics}, however, only a few synthetic methods have been shown to produce such particles and the methods tend to suffer from batch-to-batch reproducibility, with FeO being a common impurity phase that reduces the magnetization. Optimization of the synthetic processes requires rapid screening to evaluate the magnetic properties, which could be integrated into an autonomous synthesis workflow. 
\par
In this work, we present a method for generating simulated XRD data, paired with magnetization versus magnetic field for iron oxide nanoparticles. The simulated data is then used to train a Random Forest (RF) regression model \cite{breiman2001random} and a Gradient Boosting (GB) regression model \cite{friedman2001greedy} to predict the magnetization from the XRD patterns. Using the results of inference on experimental data, the models are tuned for training dataset size and hyperparameters, yielding strong predictive metrics on the five experimental datasets used for inference. We illustrate this approach schematically in Figure \ref{fig:Fig1}. Our work gives an approach that could be used to seed a ML process within a SDL before any experimental data has been collected. Additionally, our approach could be used to augment experimental data in training ML models for property prediction from XRD, where reaching the required data quantities can be challenging, depending on the synthesis method.

\section*{Simulated XRD and M-B Dataset Generation}

\begin{figure*}
    \centering
    \includegraphics[width = 1.0\textwidth]{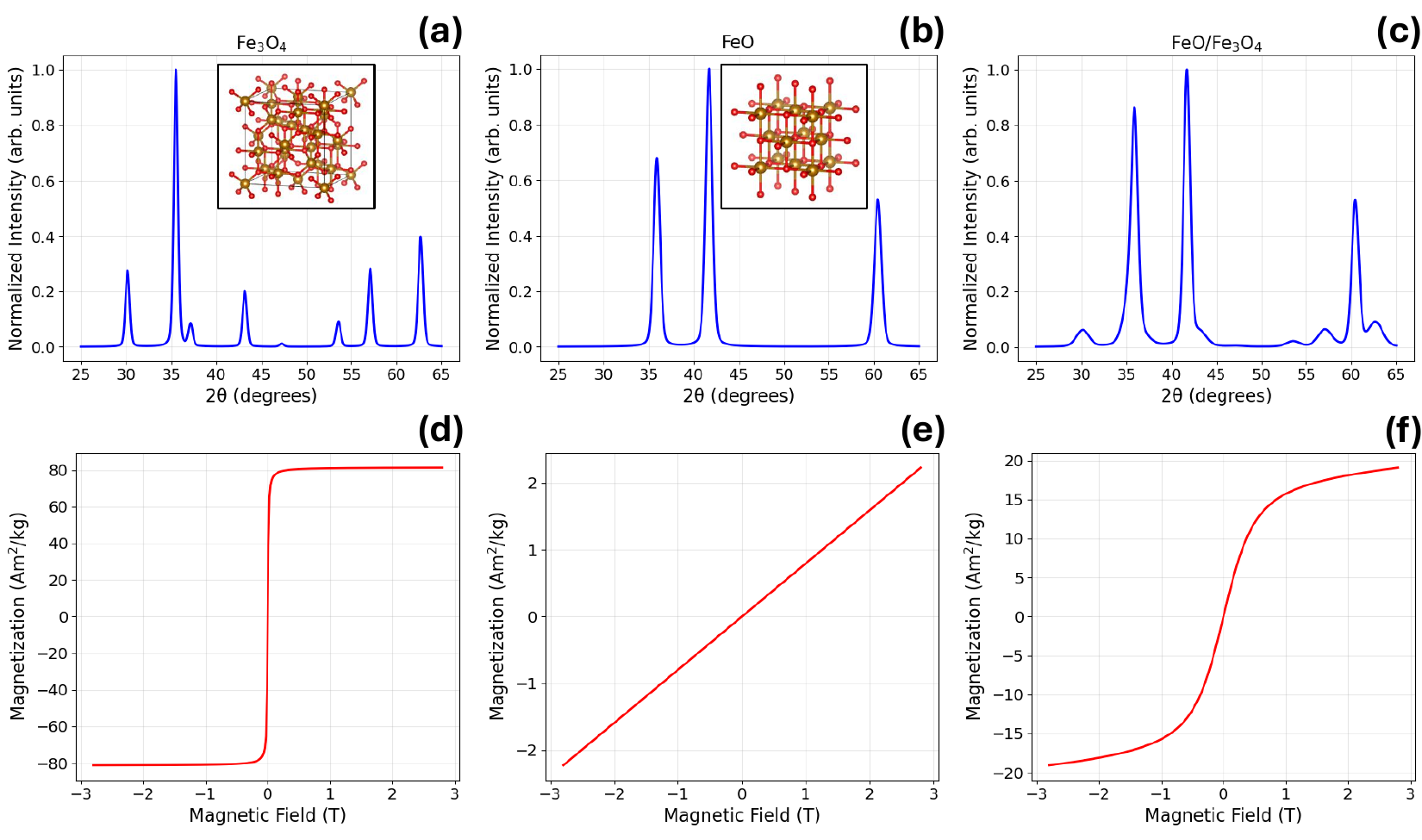}
     \caption{Possible pairs of XRD and magnetization (M) vs. magnetic field (B) for iron oxide nanoparticles that could be generated in a given dataset of size N. XRD and magnetization vs. magnetic field for Fe$_3$O$_4$ (a,c), FeO (b,d), and mixture of FeO/Fe$_3$O$_4$ (e,f).}
    \label{fig:Fig2}
\end{figure*}

The simulated XRD intensity, \( I(2\theta) \), for phase mixtures of Fe$_3$O$_4$ and FeO is calculated by summing pseudo-Voigt profiles incorporating Scherrer \cite{langford1978scherrer} and instrumental broadening:
\begin{multline}
I(2\theta) = \sum_{j=\text{phases}} f_j \sum_{i=\text{peaks}} I_{0,i,j} 
[ \eta L(2\theta;2\theta_{0,i,j}, \gamma_{i,j}) \\+ (1-\eta) G(2\theta;2\theta_{0,i,j}, \sigma_{i,j}) ].
\end{multline}
Here $f_j$ is the volume fraction of phase $j$, with $\sum f_j = 1$, and $I_{0,i,j}$ and $2\theta_{0,i,j}$ are the intensity and position of peak $i$ from phase $j$, respectively, obtained from CIF data. The CIF files are obtained from the crystallography open database (COD) \cite{Vaitkus2023, Merkys2023, Vaitkus2021, Quiros2018, Merkys2016, Grazulis2015, Grazulis2012, Grazulis2009, Downs2003} and correspond to data entries 1010369 (Fe$_3$O$_4$) \cite{montoro1938miscibilita} and 1011169 (FeO).\cite{jette1933x} $\eta = 0.3$ is the mixing parameter of the Lorentzian and Gaussian. The Lorentzian ($L$) and Gaussian ($G$) profiles are defined as:
\begin{align}
L(2\theta;2\theta_0,\gamma) &= \frac{1}{\pi}\frac{\gamma}{(2\theta-2\theta_0)^2+\gamma^2}, \\
G(2\theta;2\theta_0,\sigma) &= \frac{1}{\sigma\sqrt{2\pi}}\exp\left(-\frac{(2\theta-2\theta_0)^2}{2\sigma^2}\right),
\end{align}
where Gaussian ($\sigma$) and Lorentzian ($\gamma$) parameters are related to the total full width at half maximum (FWHM) by:
\begin{equation}
\sigma = \frac{\beta_\text{total}}{2\sqrt{2\ln(2)}},\quad\gamma=\frac{\beta_\text{total}}{2},
\end{equation}
where the peak FWHM are derived from combined Scherrer and instrumental broadening:
\begin{align}
\beta_\text{size} &= \frac{K\lambda}{D\cos\theta},\\[6pt]
\beta_\text{total} &= \sqrt{\beta_\text{size}^2+\beta_\text{inst}^2},
\end{align}
where the shape factor is $K = 0.9$, wavelength is $\lambda = 1.54184\,\text{\AA}$ (Cu K$_\alpha$), the instrumental broadening is $\beta_\text{inst} = 0.02^\circ$, and $D$ is crystallite size in Angstroms. \cite{cullity2001xrd} Lastly, each simulated XRD is normalized as follows:
\begin{equation}
I_\text{norm}(2\theta) = \frac{I(2\theta)}{\max[I(2\theta)]}.
\end{equation}
The magnetization vs. magnetic field (B) is modeled by combining superparamagnetic Langevin behavior and paramagnetic FeO as follows: 
\begin{equation}
M(B) = f_{\text{Fe}_3\text{O}_4} M_s(D)\,\mathcal{L}\left(\frac{\mu_\text{eff}B}{k_B T}\right) + (1 - f_{\text{Fe}_3\text{O}_4})\frac{\chi_\text{FeO}}{\mu_0}B,
\end{equation}
where $f_{\text{Fe}_3\text{O}_4}$ is the Fe$_3$O$_4$ phase fraction, and $M_s(D)$ is the size-dependent saturation magnetization considering a spin canted/ defect surface layer (t), defined as follows:
\begin{equation}
    M_s(D)=M_{s,\text{bulk}}\left(\frac{D-2t}{D}\right)^3.
\end{equation}
The Langevin function ($\mathcal{L}(x)$) is defined as:
\begin{equation}
    \mathcal{L}(x) = \coth\left(x\right)-\frac{1}{\left(x\right)}
\end{equation}
with $x = \frac{\mu_\text{eff}B}{k_B T}$, and the effective magnetic moment per nanoparticle calculated as $\mu_\text{eff} = M_s(D)\rho V$, with units of Am$^2$. \cite{fonseca2002superparamagnetism, usov2020equilibrium} Computationally, we define the magnetic field (B) as $B = \mu_0H$ \cite{griffiths2017electrodynamics} for consistency with the experimental measurements where only the applied magnetic field ($\mu_0H)$ is measured. Lastly, the susceptibility of FeO is given as $\chi_\text{FeO}=1\times10^{-6}\,\text{m}^3/\text{kg}$. The various constants are given in Table S1. Figure \ref{fig:Fig2} shows possible pairs of XRD and magnetization vs. magnetic field datasets that could be generated within our simulation framework. Figure S1 shows how the magnetization at 2.8 T, defined as the maximum (max) magnetization, depends on crystallite size and phase fraction of Fe$_3$O$_4$ for a dataset of size N = 1000. For a given dataset, the crystallite sizes of both Fe$_3$O$_4$ and FeO are independently sampled from a uniform random distribution between 3\,nm and 50\,nm, and the Fe$_3$O$_4$ phase fraction is sampled from a uniform random distribution between 0 and 1. A single dataset with a random seed of ``42'' is the primary focus of the main text and supporting information; however, we consider the effect of different random seeds on how the models' predictions improve with the change of the ``max\_features'' hyperparameter, as will be discussed. 

\section*{Model Training, Feature Analysis, and Validation on Experimental Data}

\begin{figure*}
    \centering
    \includegraphics[width = 0.7\textwidth]{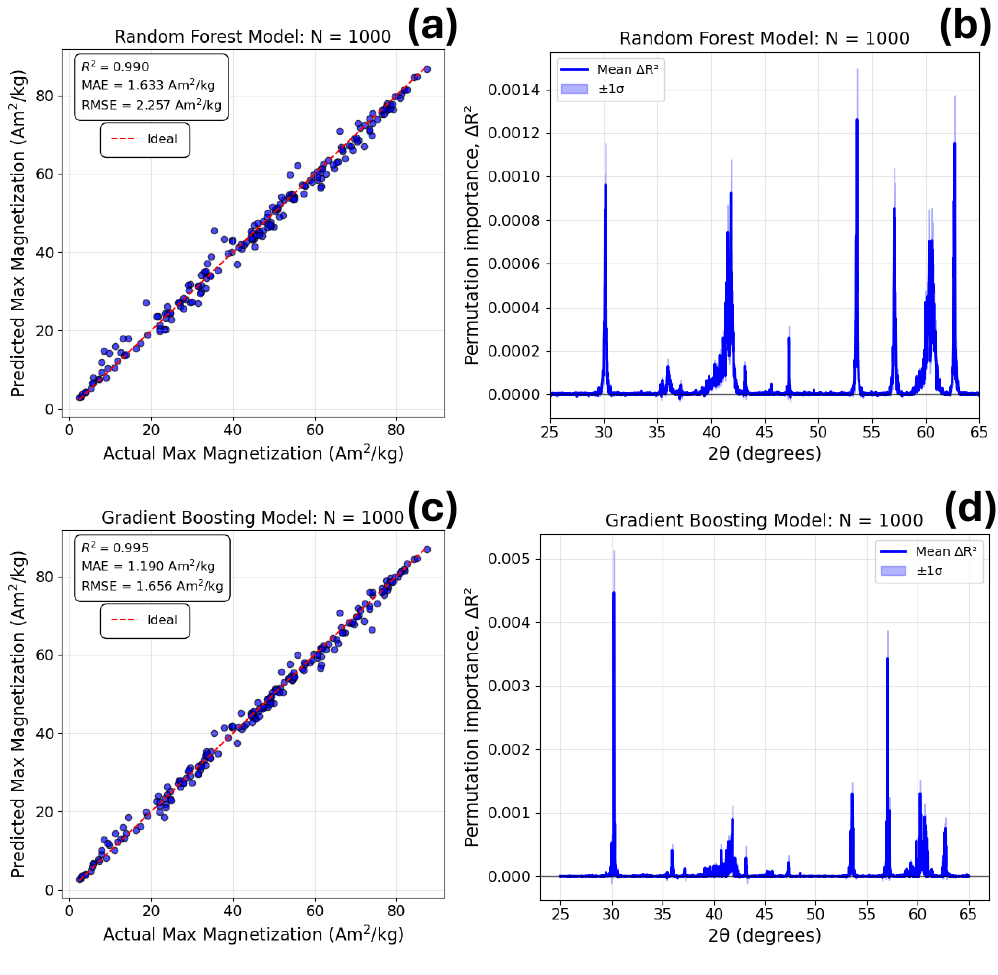}
     \caption{RF regression model predictions of max magnetization vs. true max magnetization value on 20 \% holdout data on N = 1000 dataset with R$^2$, MAE, and RMSE metrics, ideal line to indicate 1:1 correspondence (a), corresponding RF feature importance analysis, looking at change in $R^2$ with 2$\theta$ (b). GB regression model predictions of max magnetization vs. true max magnetization value on 20 \% holdout data on N = 1000 dataset with $R^2$, MAE, and RMSE metrics, ideal line to indicate 1:1 correspondence (c), corresponding GB feature importance analysis, looking at change in R$^2$ with 2$\theta$ (d).}
    \label{fig:Fig3}
\end{figure*}

Using the data simulation model, a dataset of size N can be generated where each entry contains: a 1D array of intensity values on an evenly spaced 2$\theta$ grid of 2000 points, a full 1D array of magnetization values on a fixed magnetic field grid from -2.8 T to 2.8 T, a value of the crystallite size of Fe$_3$O$_4$ and FeO, the phase fraction of Fe$_3$O$_4$, the max magnetization (magnetization at 2.8 T). With this data set, an ML model can be trained to predict the max magnetization or full magnetization curves. For each trained model, we holdout 20 \% of the data for inference on unseen simulated data. Figure \ref{fig:Fig3} (a,c) shows plots of predicted vs. true max magnetization using the holdout data along with prediction metrics, $R^2$, mean average error (MAE), root mean square error (RMSE), on a dataset of N = 1000 for the trained RF and GB models. Both models (RF and GB) show strong prediction metrics for the prediction of max magnetization, which is unsurprising when testing on correlated simulated data. The model hyperparameters are summarized in Table S2. 
As will be discussed, the optimization of these models was performed on the experimental datasets, since the correlated nature of the simulated data and the size of the datasets tends to always produce relatively good $R^2$ ($\gtrsim$ 0.8) for a sufficiently large dataset $\ge$ N = 20, likely subject to the random seed used in data generation. To analyze the importance of each feature which are the intensity values at a given 2$\theta$, we perform permutations over the intensity values and look at the change in the R$^2$ value, Eq. 11. In essence one set of intensity values is shuffled at a certain 2$\theta$ of the holdout set, it is then repeated n = 30 times and averaged, and further repeated for each set of intensity values at each 2$\theta$.
\begin{equation}
\text{Importance}(i) =R^2_{\text{baseline}} - \frac{1}{n} \sum_{j=1}^{n} R^2_{\text{perm.},j}(i)
\end{equation}
 Figure \ref{fig:Fig3} (b,d) show the plots of $\Delta R^2$ as a function of 2$\theta$ for the RF and GB models. Since the regression is on simulated data, the change in $R^2$ is relatively small; however, it gives us insight into which peaks in the XRD pattern of the iron oxides are important for the prediction of the max magnetization value. Both models strongly consider the peak at about 30$^o$ associated with the Fe$_3$O$_4$ phase. Likely because it's one of the highest intensity peaks of the Fe$_3$O$_4$ phase, that is distinct from peaks associated with FeO, which makes it an easy identifier of Fe$_3$O$_4$ even with a large amount of FeO that may obscure the other main peaks of Fe$_3$O$_4$. However, the RF model in comparison to the GB model shows a relatively greater change in R$^2$ with a larger number of peaks, possibly indicating a better generalized learning to look at a more diverse feature space when evaluating a pattern for the prediction of max magnetization.
\par
\begin{figure*} 
    \centering
    \includegraphics[width = 1.0\textwidth]{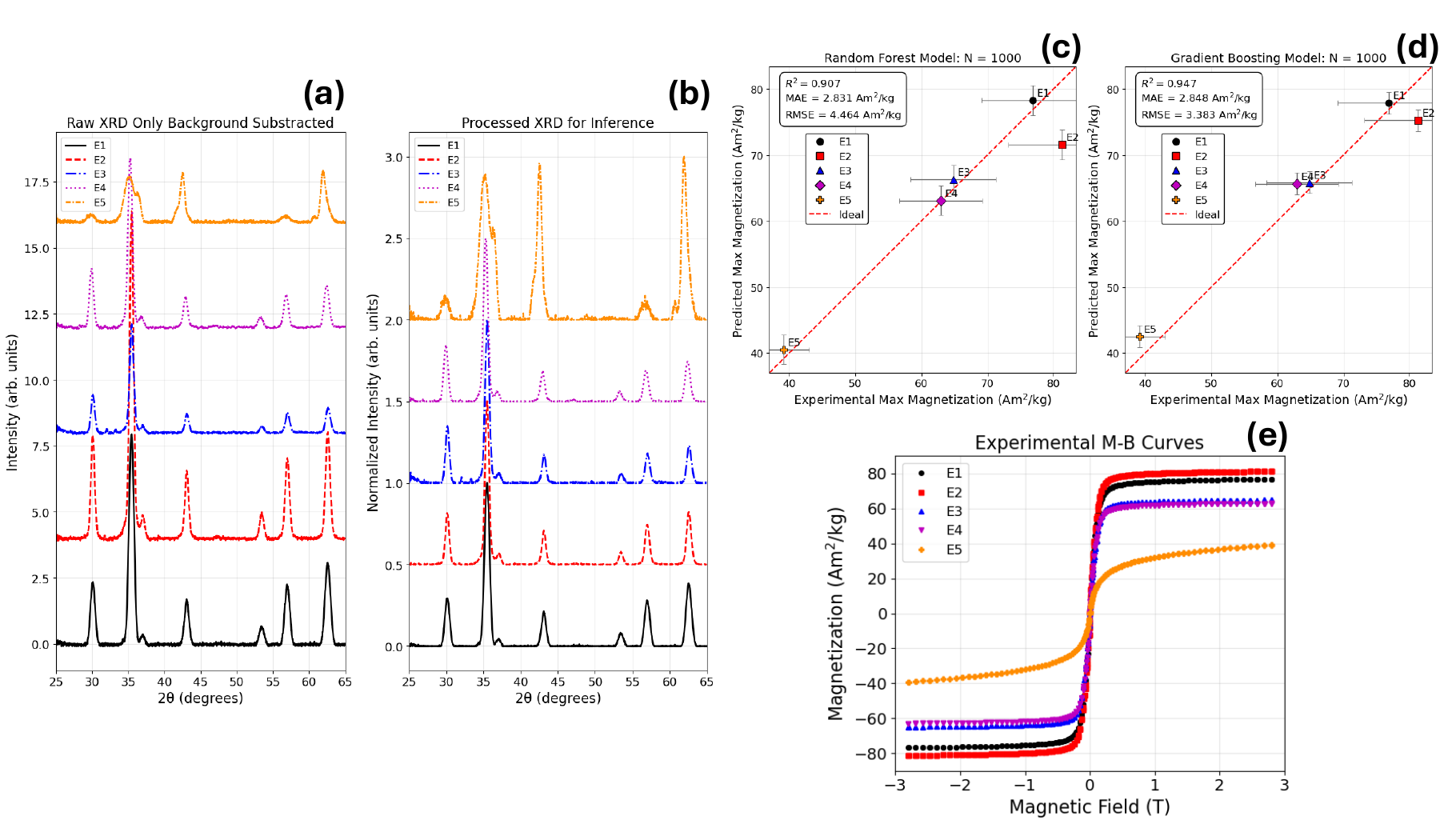}
     \caption{Raw experimental XRD (background subtraction only) data of five iron oxide nanoparticle samples (a), experimental XRD after pre-processing for inference (b), RF model prediction of max magnetization vs. experimentally measured max magnetization (c), GB model prediction of max magnetization vs. experimentally measured max magnetization (d), in both cases the error in prediction is taken as the RMSE of the models prediction of holdout simulated data and the error in experimental max magnetization is estimated as 10 \% of the sample mass. Experimental measurements of magnetization vs. magnetic field, the max magnetization corresponds to the value at maximum magnetic field about 2.8 T (e).}
    \label{fig:Fig4}
\end{figure*}

To validate and improve the training of the models (hyperparameter optimization) and the general approach of generating simulated XRD and magnetization curves for iron oxide nanoparticles, we synthesized a series of iron oxide nanoparticles and characterized the samples by XRD and vibrating sample magnetometry (VSM). The five experimental samples consist of one mixed phase sample FeO/Fe$_3$O$_4$/$\gamma$-Fe$_2$O$_3$, three nearly pure phase Fe$_3$O$_4$/$\gamma$-Fe$_2$O$_3$, one of which was doped with Zn (likely Zn$_x$Fe$_{3-x}$O$_4$), and finally a majority Fe$_3$O$_4$/$\gamma$-Fe$_2$O$_3$ phase sample with an impurity phase not represented in the simulated data. We note both Fe$_3$O$_4$ and $\gamma$-Fe$_2$O$_3$ as the inverse spinel phase in all the samples since these two phases are not easily distinguishable by XRD, and often resides on the surface of Fe$_3$O$_4$. \cite{baaziz2014magnetic, sharifi2018determining} Inference on the five experimental XRD patterns requires pre-processing to attempt to match as closely as possible to the simulated data that the models were trained on. First, background subtraction was performed on the raw XRD patterns using the Bruker Diffract.EVA software. The data is then cut to the same 2$\theta$ range as the simulated (25$^o$ to 65$^o$), negative values are set to zero, and the patterns are normalized to 1. Next, the experimental data is resampled by linear interpolation onto the same uniform 2$\theta$ grid as the simulated patterns. Lastly, the data is normalized to 1 again. All pre-processing, except background subtraction, is performed within the inference script. Figure \ref{fig:Fig4} shows the raw XRD with only background subtraction (a) and after pre-processing (b). E1, E2, and E4 are the nearly pure phase (Fe$_3$O$_4$/$\gamma$-Fe$_2$O$_3$) samples (E1 was doped with Zn, likely Zn$_x$Fe$_{3-x}$O$_4$), E3 has a majority Fe$_3$O$_4$/$\gamma$-Fe$_2$O$_3$ with an impurity phase not included in the simulated training data, and E5 is a mixture of FeO/Fe$_3$O$_4$/$\gamma$-Fe$_2$O$_3$. Figure \ref{fig:Fig4} (c, d) shows the prediction of max magnetization on the five experimental XRD measurements using the RF and GB models after model tuning, which improved prediction performance. Both models show surprisingly promising prediction metrics, both achieve greater than 0.9 R$^2$, MAE at approximately 2.8 Am$^2$/kg, and RMSE from about (3.4 to 4.5) Am$^2$/kg. For example, a high crystal quality iron oxide sample might have a high field magnetization in the range of 70 Am$^2$/kg to 85 Am$^2$/kg, depending on size, which would mean a deviation of about 4 Am$^2$/kg would only be about a 6 \% to 5 \% change from the experimental value, respectively. The experimental magnetization vs. magnetic field curves from which the experimental max magnetization was obtained (value at approximately 2.8 T) are shown in Figure \ref{fig:Fig4} (e). 
\par
As mentioned, due to the linked nature of the simulated data, model tuning could really only be performed using experimental data, since minimal changes in prediction metrics are observed when inference is performed on the holdout simulated data. Due to the numerous potential parameters between the hyperparameters of each model and the various parameters related to how the simulated data are generated, we impose limited tuning conditions to avoid biasing the model too strongly toward the small experimental dataset. For dataset size, we find that both RF and GB models using a dataset of N = 1000 seem to be an ideal simulated data quantity, with RF changing minimally between N = 1000 and N = 10000 (Figure S2 (a)), and the GB model prediction metrics generally get worse between N = 1000 and N = 10000 (Figure S2 (b)). Regarding hyperparameters, we explored ``max\_features'' (a parameter of both RF and GB models), changing it from 0.3 to ``sqrt'' corresponding to 600 to about 44.7, for the 2000 features in the simulated XRD data. In this dataset (N = 1000), the effect was similar across the RF and GB models. For the RF model, the prediction metrics improved from an $R^2$ of 0.877 to 0.907, and MAE and RMSE decreased from about 4.0 Am$^2$/kg to 2.8  Am$^2$/kg, and 5.2  Am$^2$/kg to 4.5  Am$^2$/kg, respectively, Figure S3 (a, c). Similarly, for the GB model, $R^2$ improved from 0.916 to 0.947, and MAE and RMSE decreased from about 3.4 Am$^2$/kg to 2.8 Am$^2$/kg, and 4.3 Am$^2$/kg to 3.4 Am$^2$/kg, respectively, Figure S4 (a, c). Considering the feature analysis of holdout simulated data for the RF model, it's clear that decreasing the ``max\_features'' increases the importance of more peaks across a given pattern, Figure S3 (b, d). Using a small ``max\_features'' parameter seems important in training RF models for high-dimensional correlated datasets like XRD. The change forces each tree to consider only a small random subset of 2$\theta$ values, thereby reducing overfitting and, in this case, likely improving generalization. It seems critical for the forest to learn subtle features like minor peaks and shoulders, which can strongly correlate to the magnetic properties. However, for the GB model, despite showing similar improvements in prediction metrics, the feature analysis reveals similar feature importance with this change in ``max\_features'', as shown in Figure S4 (b, d). To explore this observation more generally, we examine the effect of the random seed on the generation of the simulated datasets. Thus far, we have considered only a single random seed, and changing the seed can slightly alter the parameter distribution for a given dataset of size N. We consider four different random seeds for N = 1000 (including the one used in the main text) for both ``max\_features''. Table S3 summarizes the prediction metrics across different random seeds for the RF and GB models with ``max\_features'' set to 0.3 and ``sqrt''. By considering the mean prediction metrics of all four seeds, we observe that both models show improvements; however, improvements of the RF model are significantly greater when using ``max\_features'' of ``sqrt'' instead of 0.3, with $R^2$ improving by about a factor of 2, and MAE improving by about a factor of 3, relative to the GB model's improvements ($\Delta R^2_{\mathrm{RF}} \approx 0.108$ vs.\ $\Delta R^2_{\mathrm{GB}} \approx 0.054$ and $\Delta MAE_{\mathrm{RF}} \approx 2.8$ Am$^2$/kg vs.\ $\Delta MAE_{\mathrm{GB}} \approx 0.87$ Am$^2$/kg, Table S3). The difference in how RF and GB models respond to changes in ``max\_features'' likely stems from RF's greater sensitivity to the number of features tried at each split compared to GB. We speculate that, although the GB model slightly outperforms the RF model on our small experimental dataset, the RF model's more diverse feature importance, as shown in Figure S3 (c), will offer better generalization when considering additional experimental data. Since one of the primary goals of our work was to develop an ML model for high-throughput screening of iron oxide nanoparticle samples, we investigated the effect of XRD measurement time on the prediction of magnetization. Figure S5 (a) shows three different XRD measurements of sample E2, taken with three different measurement times, changing by an order of magnitude. Figure S5 (b, c) shows that for the best performing models, hyperparameters in Table S2, the accuracy of the prediction doesn't significantly change between a measurement of 2052 s ($\approx$ 34 min.)  and 20520 s ($\approx$ 6 h). However, there is a significant effect with background subtraction. Our result is extremely promising for screening large quantities of samples (30 to 50) in a relatively short time-frame, and we show the importance of background subtraction when pre-processing the datasets. 
\begin{figure*}
    \centering
    \includegraphics[width = 1.0\textwidth]{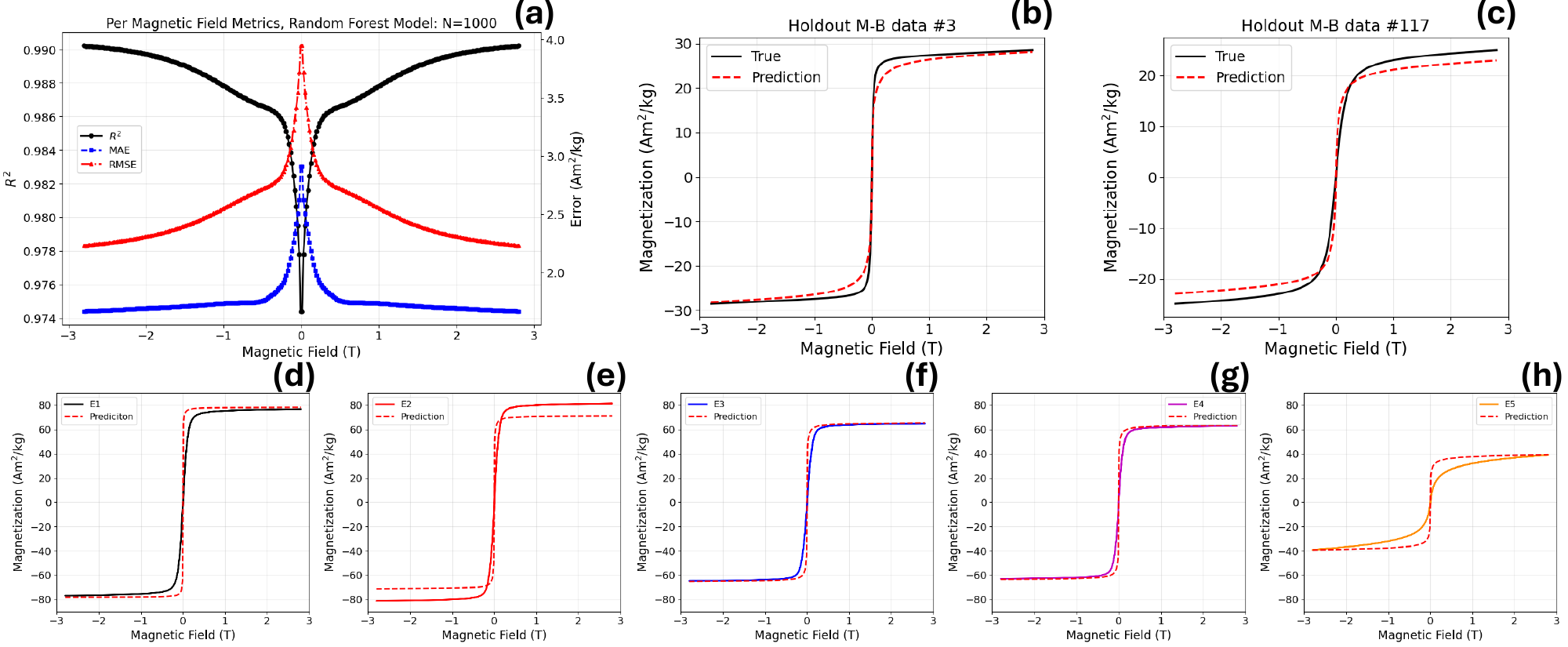}
     \caption{Prediction metrics for the RF model at each magnetic field value for holdout simulated data (a), full M-B curve RF prediction on holdout simulated data (b, c), and full M-B curve RF prediction compared to the experimentally measured M-B curves for the five iron oxide nanoparticle samples (d-h).}
    \label{fig:Fig5}
\end{figure*}
\par
So far, we have only considered training the models on XRD patterns correlated to a single magnetization value at the maximum magnetic field. However, as mentioned, our simulation model can generate full magnetization vs. magnetic field curves, and we can train a model to go from 1D XRD intensity to 1D magnetization. With the 1D magnetization being on a fixed grid of magnetic field values between -2.8 T to 2.8 T with a uniform spacing of 300 points. For sample screening, prediction of the max magnetization value is most critical, but the prediction of the full curve could produce other useful metrics like dM/dB. The first derivative of the Langevin-like M-B curve has been used as a metric to evaluate potential imaging resolution for MPI \cite{goodwill2011multidimensional}, since it can roughly correlate to the switching of the magnetization when the nanoparticles are subjected to an AC magnetic field. Figure \ref{fig:Fig5} (a), shows the prediction metrics as a function of magnetic field for holdout simulated data using the previously optimized RF model, hyperparameters in Table S2, with N = 1000. Interestingly, we see that the model performs the best at high fields and decreases performance for magnetic fields around zero. Figure \ref{fig:Fig5} (b, c) shows examples of predicted and true (simulated) magnetization vs. magnetic field curves. Performing inference on the five experimental XRD patterns, the predicted curves are plotted with the experimental measurements in Figure \ref{fig:Fig5} (d-h). Testing on the experimental data clearly shows a similar trend to the per-magnetic-field metrics on holdout simulated data, with the high-field magnetization being extremely close. However, the model fails to accurately predict the overall shape of the curves.

\section*{Discussion and Conclusion}

The idea of closed-loop autonomous material fabrication systems, or SDL, is continuously gaining traction as the future of material synthesis and design. These systems require ML models to predict material properties, guiding the next step in the synthesis optimization process. In this work, we demonstrate how a simulated dataset generated from simple physical models can be used to train both RF and GB models to accurately predict high-field magnetization of iron oxide nanoparticles from experimental XRD data. At its core, the physical model describes how magnetization changes based on the crystallite size and the volume fraction of the two phases. The RF and GB models can then learn the features that correlate to these quantities and make reasonable predictions of the magnetization. While these two characteristics lead to good predictions of the high-field magnetization, the RF model generally fails to accurately predict the full magnetization curves. This is likely in part because the magnetic models are too simple to capture many real physical features, such as the finite coercivity that can be observed in iron oxide nanoparticles or inter-particle interactions, more aptly captured by micromagnetics. Although this failing holds for both simulated and experimental data, meaning additional tuning could improve predictions of the full magnetization curve, at least on holdout simulated data. However, we show that training models purely on simulated data and minimal tuning on a small set of experimental data can yield reasonable predictions on that dataset. Our finding is key, since the dataset size to train RF and GB models is typically unobtainable on a reasonable timescale for many synthetic processes. The ability to use simulated data to train the first iterative model that can make accurate predictions of a key property will be important to ``seed" a closed-loop system, after which experimental data can be integrated into the model to improve the optimization of the synthesis process. The alternative non-ML approach for making an estimate of the magnetization from XRD would be through using Rietveld refinement to determine the phase fractions and crystallite size, which could be used in the magnetic model to determine the magnetization curve. However, Rietveld refinement is often time-consuming, difficult to automate, requires specialized software, and complete knowledge of the possible phases present in the sample. \cite{tamura2022automatic}. For these reasons, an ML approach is preferable for integration into an SDL or for high-throughput characterization. Additionally, the workflow could be extended to other material systems, incorporating more complex information from XRD, such as multiple impurity phases, texture, strain, and experimental noise. This could then be coupled with a multi-scale approach using DFT-derived intrinsic magnetic parameters, which could be used as input into a micromagnetic simulation to capture more realistic magnetic hysteresis. \cite{kazantseva2008towards} In closing, we have demonstrated that a simple physics-informed magnetic model coupled with corresponding XRD patterns with a shared set of parameters can be used to generate simulated data in large quantities (N = 1000), far exceeding what could be done experimentally, and then be used to make accurate predictions on experimental XRD of the high-field magnetization of iron oxide nanoparticles.

\section*{Experimental Methods}

XRD measurements were performed with a Bruker D8 Discover using Cu K$_{\alpha}$ X-rays, the samples were drop casted from a colloidal iron oxides in hexanes on to a zero-diffract Si disc. Magnetic measurements were performed on a Quantum Design (QD) VersaLab. The samples were dried from a colloidal state to a nanocrystalline powder and loaded into a QD powder capsule. The magnetization of all samples were measured at 300 K between -2.8 T and 2.8 T (initial curve and full hysteresis).

\section*{Supporting Information}

The supporting information (SI) includes a table of magnetic parameters used in the models to generate simulated data, and a table of hyperparameters used in the validated and tuned RF and GB models. The SI also contains figures showing the relation between max magnetization and shared parameters in the simulated datasets, the effect of simulated dataset size on the performance metrics for inference on the experimental dataset, the effect of changing the ``max\_features" hyperparameter for training the RF and GB models, a table that shows the prediction metrics for different random seeds used to generate the simulated datasets, and the effect of the prediction of max magnetization based on XRD data measurement time with and without background subtraction.

\section*{Code and Data Availability}

Simulated data generation, model training, and inference scripts are available at \href{https://github.com/Fabel88/XRD-to-magnetic-properties}{GitHub}. XRD and magnetic measurement data are provided on \href{https://github.com/Fabel88/XRD-to-magnetic-properties}{GitHub} for the purpose of running the inference scripts.

\section*{Disclaimer}

Certain commercial equipment, instruments, software, or materials are identified in this paper in order to specify the experimental procedure adequately. Such identifications are not intended to imply recommendation or endorsement by the National Institute of Standards and Technology (NIST), nor it is intended to imply that the materials or equipment identified are necessarily the best available for the purpose.

The views expressed in this article are those of the authors and do not reflect the official policy or position of the U.S. Naval Academy, Department of the Navy, the Department of Defense, or the U.S. Government.

\clearpage
\label{References}
\bibliographystyle{apsrev4-1}
\bibliography{bib}

@article{lunt2024modular,
  title={Modular, multi-robot integration of laboratories: an autonomous workflow for solid-state chemistry},
  author={Lunt, Amy M and Fakhruldeen, Hatem and Pizzuto, Gabriella and Longley, Louis and White, Alexander and Rankin, Nicola and Clowes, Rob and Alston, Ben and Gigli, Lucia and Day, Graeme M and others},
  journal={Chemical Science},
  volume={15},
  number={7},
  pages={2456--2463},
  year={2024},
  publisher={Royal Society of Chemistry}
}

@article{vriza2023self,
  title={Self-driving laboratory for polymer electronics},
  author={Vriza, Aikaterini and Chan, Henry and Xu, Jie},
  journal={Chemistry of Materials},
  volume={35},
  number={8},
  pages={3046--3056},
  year={2023},
  publisher={ACS Publications}
}

@article{macleod2020self,
  title={Self-driving laboratory for accelerated discovery of thin-film materials},
  author={MacLeod, Benjamin P and Parlane, Fraser GL and Morrissey, Thomas D and H{\"a}se, Florian and Roch, Lo{\"\i}c M and Dettelbach, Kevan E and Moreira, Raphaell and Yunker, Lars PE and Rooney, Michael B and Deeth, Joseph R and others},
  journal={Science Advances},
  volume={6},
  number={20},
  pages={eaaz8867},
  year={2020},
  publisher={American Association for the Advancement of Science}
}

@article{breiman2001random,
  title={Random forests},
  author={Breiman, Leo},
  journal={Machine learning},
  volume={45},
  number={1},
  pages={5--32},
  year={2001},
  publisher={Springer}
}

@article{friedman2001greedy,
  title={Greedy function approximation: a gradient boosting machine},
  author={Friedman, Jerome H},
  journal={Annals of statistics},
  pages={1189--1232},
  year={2001},
  publisher={JSTOR}
}

@article{szymanski2023autonomous,
  title={An autonomous laboratory for the accelerated synthesis of novel materials},
  author={Szymanski, Nathan J and Rendy, Bernardus and Fei, Yuxing and Kumar, Rishi E and He, Tanjin and Milsted, David and McDermott, Matthew J and Gallant, Max and Cubuk, Ekin Dogus and Merchant, Amil and others},
  journal={Nature},
  volume={624},
  number={7990},
  pages={86--91},
  year={2023},
  publisher={Nature Publishing Group UK London}
}

@article{bhandari2024accurate,
  title={Accurate machine-learning predictions of coercivity in high-performance permanent magnets},
  author={Bhandari, Churna and Nop, Gavin N and Smith, Jonathan DH and Paudyal, Durga},
  journal={Physical Review Applied},
  volume={22},
  number={2},
  pages={024046},
  year={2024},
  publisher={APS}
}

@article{qiao2023performance,
  title={Performance prediction models for sintered NdFeB using machine learning methods and interpretable studies},
  author={Qiao, Zuqiang and Dong, Shengzhi and Li, Qing and Lu, Xiangming and Chen, Renjie and Guo, Shuai and Yan, Aru and Li, Wei},
  journal={Journal of Alloys and Compounds},
  volume={963},
  pages={171250},
  year={2023},
  publisher={Elsevier}
}

@article{kovacs2023physics,
  title={Physics-informed machine learning combining experiment and simulation for the design of neodymium-iron-boron permanent magnets with reduced critical-elements content},
  author={Kovacs, Alexander and Fischbacher, Johann and Oezelt, Harald and Kornell, Alexander and Ali, Qais and Gusenbauer, Markus and Yano, Masao and Sakuma, Noritsugu and Kinoshita, Akihito and Shoji, Tetsuya and others},
  journal={Frontiers in Materials},
  volume={9},
  pages={1094055},
  year={2023},
  publisher={Frontiers Media SA}
}

@article{liu2022magnetic,
  title={The magnetic properties prediction and composition design of La-Co substitution Sr-hexaferrite based on high-through experiments and machine learning},
  author={Liu, Ruoshui and Wang, Lichen and Xu, Zhiyi and Qin, Ciyu and Li, Ziyu and Yu, Xiang and Liu, Dan and Gong, Huayang and Zhao, Tongyun and Sun, Jirong and others},
  journal={Materials Today Communications},
  volume={32},
  pages={103996},
  year={2022},
  publisher={Elsevier}
}

@article{dengina2022machine,
  title={Machine learning approach for evaluation of nanodefects and magnetic anisotropy in FePt granular films},
  author={Dengina, E and Bolyachkin, A and Sepehri-Amin, H and Hono, K},
  journal={Scripta Materialia},
  volume={218},
  pages={114797},
  year={2022},
  publisher={Elsevier}
}

@article{xia2022accelerating,
  title={Accelerating the discovery of novel magnetic materials using machine learning--guided adaptive feedback},
  author={Xia, Weiyi and Sakurai, Masahiro and Balasubramanian, Balamurugan and Liao, Timothy and Wang, Renhai and Zhang, Chao and Sun, Huaijun and Ho, Kai-Ming and Chelikowsky, James R and Sellmyer, David J and others},
  journal={Proceedings of the National Academy of Sciences},
  volume={119},
  number={47},
  pages={e2204485119},
  year={2022},
  publisher={National Academy of Sciences}
}

@article{abel2024strongly,
  title={Strongly Interacting Nanoferrites for Magnetic Particle Imaging and Spatially Resolved Thermometry},
  author={Abel, Frank M and Correa, Eduardo L and Bui, Thinh Q and Biacchi, Adam J and Donahue, Michael J and Merritt, Mia T and Seppala, Jonathan E and Woods, Solomon I and Hight Walker, Angela R and Dennis, Cindi L},
  journal={ACS Applied Materials \& Interfaces},
  volume={16},
  number={40},
  pages={54328--54343},
  year={2024},
  publisher={ACS Publications}
}

@article{tay2021superferromagnetic,
  title={Superferromagnetic nanoparticles enable order-of-magnitude resolution \& sensitivity gain in magnetic particle imaging},
  author={Tay, Zhi Wei and Savliwala, Shehaab and Hensley, Daniel W and Fung, KL Barry and Colson, Caylin and Fellows, Benjamin D and Zhou, Xinyi and Huynh, Quincy and Lu, Yao and Zheng, Bo and others},
  journal={Small methods},
  volume={5},
  number={11},
  pages={2100796},
  year={2021},
  publisher={Wiley Online Library}
}

@article{fung2023first,
  title={First superferromagnetic remanence characterization and scan optimization for super-resolution magnetic particle imaging},
  author={Fung, KL Barry and Colson, Caylin and Bryan, Jacob and Saayujya, Chinmoy and Mokkarala-Lopez, Javier and Hartley, Allison and Yousuf, Khadija and Kuo, Renesmee and Lu, Yao and Fellows, Benjamin D and others},
  journal={Nano letters},
  volume={23},
  number={5},
  pages={1717--1725},
  year={2023},
  publisher={ACS Publications}
}

@article{bui2023harmonic,
  title={Harmonic dependence of thermal magnetic particle imaging},
  author={Bui, Thinh Q and Henn, Mark-Alexander and Tew, Weston L and Catterton, Megan A and Woods, Solomon I},
  journal={Scientific Reports},
  volume={13},
  number={1},
  pages={15762},
  year={2023},
  publisher={Nature Publishing Group UK London}
}

@article{abel2023thermosensitivity,
  title={Thermosensitivity through exchange coupling in ferrimagnetic/antiferromagnetic nano-objects for magnetic-based thermometry},
  author={Abel, Frank M and Correa, Eduardo L and Biacchi, Adam J and Bui, Thinh Q and Woods, Solomon I and Hight Walker, Angela R and Dennis, Cindi L},
  journal={ACS applied materials \& interfaces},
  volume={15},
  number={10},
  pages={13439--13448},
  year={2023},
  publisher={ACS Publications}
}

@article{bui2024magnetodynamics,
  title={Magnetodynamics of few-nanoparticle chains},
  author={Bui, Thinh Q and Oberdick, Samuel D and Abel, Frank M and Donahue, Michael J and Quelhas, Klaus N and Dennis, Cindi L and Cleveland, Thomas and Liu, Yanxin and Woods, Solomon I},
  journal={arXiv preprint arXiv:2408.01561},
  year={2024}
}

@article{goodwill2011multidimensional,
  title={Multidimensional x-space magnetic particle imaging},
  author={Goodwill, Patrick W and Conolly, Steven M},
  journal={IEEE transactions on medical imaging},
  volume={30},
  number={9},
  pages={1581--1590},
  year={2011},
  publisher={IEEE}
}

@article{baaziz2014magnetic,
  title={Magnetic iron oxide nanoparticles: reproducible tuning of the size and nanosized-dependent composition, defects, and spin canting},
  author={Baaziz, Walid and Pichon, Benoit P and Fleutot, Solenne and Liu, Yu and Lefevre, Christophe and Greneche, Jean-Marc and Toumi, Mohamed and Mhiri, Tahar and Begin-Colin, Sylvie},
  journal={The Journal of Physical Chemistry C},
  volume={118},
  number={7},
  pages={3795--3810},
  year={2014},
  publisher={ACS Publications}
}

@article{sharifi2018determining,
  title={Determining magnetite/maghemite composition and core--shell nanostructure from magnetization curve for iron oxide nanoparticles},
  author={Sharifi Dehsari, Hamed and Ksenofontov, Vadim and M{\"o}ller, Angela and Jakob, Gerhard and Asadi, Kamal},
  journal={The Journal of Physical Chemistry C},
  volume={122},
  number={49},
  pages={28292--28301},
  year={2018},
  publisher={ACS Publications}
}

@article{moustafa2025graph,
  title={Graph Neural Networks to Predict Coercivity of Hard Magnetic Microstructures},
  author={Moustafa, Heisam and Kovacs, Alexander and Fischbacher, Johann and Gusenbauer, Markus and Ali, Qais and Breth, Leoni and Schrefl, Thomas and Oezelt, Harald},
  journal={arXiv preprint arXiv:2506.23615},
  year={2025}
}

@article{exl2018magnetic,
  title={Magnetic microstructure machine learning analysis},
  author={Exl, Lukas and Fischbacher, Johann and Kovacs, Alexander and Oezelt, Harald and Gusenbauer, Markus and Yokota, Kazuya and Shoji, Tetsuya and Hrkac, Gino and Schrefl, Thomas},
  journal={Journal of Physics: Materials},
  volume={2},
  number={1},
  pages={014001},
  year={2018},
  publisher={IOP Publishing}
}

@article{khurshid2013synthesis,
  title={Synthesis and magnetic properties of core/shell FeO/Fe3O4 nano-octopods},
  author={Khurshid, Hafsa and Chandra, Sayan and Li, Wanfeng and Phan, MH and Hadjipanayis, GC and Mukherjee, P and Srikanth, H},
  journal={Journal of Applied Physics},
  volume={113},
  number={17},
  year={2013},
  publisher={AIP Publishing}
}

@article{castellanos2021milestone,
  title={A milestone in the chemical synthesis of Fe3O4 nanoparticles: unreported bulklike properties lead to a remarkable magnetic hyperthermia},
  author={Castellanos-Rubio, Idoia and Arriortua, Oihane and Iglesias-Rojas, Daniela and Bar{\'o}n, Ander and Rodrigo, Irati and Marcano, Lourdes and Garitaonandia, Jos{\'e} S and Orue, Iñaki and Fdez-Gubieda, M Luisa and Insausti, Maite},
  journal={Chemistry of Materials},
  volume={33},
  number={22},
  pages={8693--8704},
  year={2021},
  publisher={ACS Publications}
}

@article{sun2002size,
  title={Size-controlled synthesis of magnetite nanoparticles},
  author={Sun, Shouheng and Zeng, Hao},
  journal={Journal of the American Chemical Society},
  volume={124},
  number={28},
  pages={8204--8205},
  year={2002},
  publisher={ACS Publications}
}

@article{langford1978scherrer,
  title={Scherrer after sixty years: a survey and some new results in the determination of crystallite size},
  author={Langford, J Il and Wilson, AJC},
  journal={Applied Crystallography},
  volume={11},
  number={2},
  pages={102--113},
  year={1978},
  publisher={International Union of Crystallography}
}

@article{fonseca2002superparamagnetism,
  title={Superparamagnetism and magnetic properties of Ni nanoparticles embedded in SiO 2},
  author={Fonseca, Fabio Coral and Goya, Gerardo Fab{\'\i}an and Jardim, RF and Muccillo, R and Carreno, NLV and Longo, El and Leite, ER},
  journal={Physical review B},
  volume={66},
  number={10},
  pages={104406},
  year={2002},
  publisher={APS}
}

@article{usov2020equilibrium,
  title={Equilibrium properties of assembly of interacting superparamagnetic nanoparticles},
  author={Usov, NA and Serebryakova, ON},
  journal={Scientific reports},
  volume={10},
  number={1},
  pages={13677},
  year={2020},
  publisher={Nature Publishing Group UK London}
}

@book{griffiths2017electrodynamics,
  title     = {Introduction to Electrodynamics},
  author    = {Griffiths, David J.},
  edition   = {4},
  year      = {2017},
  publisher = {Cambridge University Press},
  address   = {Cambridge}
}

@article{kazantseva2008towards,
  title={Towards multiscale modeling of magnetic materials: Simulations of FePt},
  author={Kazantseva, Natalia and Hinzke, Denise and Nowak, Ulrich and Chantrell, Roy W and Atxitia, Unai and Chubykalo-Fesenko, Oksana},
  journal={Physical Review B—Condensed Matter and Materials Physics},
  volume={77},
  number={18},
  pages={184428},
  year={2008},
  publisher={APS}
}

@Article{Vaitkus2023,
  author    = {Vaitkus, Antanas and Merkys, Andrius and Sander, Thomas and Quirós, Miguel and Thiessen, Paul A. and Bolton, Evan E. and Gražulis, Saulius},
  title     = {A workflow for deriving chemical entities from crystallographic data and its application to the {C}rystallography {O}pen {D}atabase},
  journal   = {Journal of Cheminformatics},
  year      = {2023},
  volume    = {15},
  number    = {1},
  month     = {Dec},
  doi       = {10.1186/s13321-023-00780-2},
  url       = {https://doi.org/10.1186/s13321-023-00780-2},
  publisher = {Springer Science and Business Media LLC},
}

@Article{Merkys2023,
  author    = {Merkys, Andrius and Vaitkus, Antanas and Grybauskas, Algirdas and Konovalovas, Aleksandras and Quir{\'{o}}s, Miguel and Gra{\v{z}}ulis, Saulius},
  title     = {Graph isomorphism-based algorithm for cross-checking chemical and crystallographic descriptions},
  volume    = {15},
  url       = {https://doi.org/10.1186/s13321-023-00692-1},
  doi       = {10.1186/s13321-023-00692-1},
  number    = {1},
  journal   = {Journal of Cheminformatics},
  publisher = {Springer Science and Business Media LLC},
  year      = {2023},
  month     = {Feb},
}

@Article{Vaitkus2021,
  author   = {Vaitkus, Antanas and Merkys, Andrius and Gražulis, Saulius},
  title    = {Validation of the {C}rystallography {O}pen {D}atabase using the {C}rystallographic {I}nformation {F}ramework},
  journal  = {Journal of Applied Crystallography},
  year     = {2021},
  volume   = {54},
  number   = {2},
  pages    = {661--672},
  month    = {Apr},
  doi      = {10.1107/S1600576720016532},
  url      = {https://doi.org/10.1107/S1600576720016532},
}

@Article{Quiros2018,
  author    = {Miguel Quir{\'{o}}s and Saulius Gra{\v{z}}ulis and Saul{\.{e}} Girdzijauskait{\.{e}} and Andrius Merkys and Antanas Vaitkus},
  title     = {Using {SMILES} strings for the description of chemical connectivity in the {C}rystallography {O}pen {D}atabase},
  journal   = {Journal of Cheminformatics},
  year      = {2018},
  volume    = {10},
  number    = {1},
  month     = {May},
  doi       = {10.1186/s13321-018-0279-6},
  publisher = {Springer Nature},
}

@ARTICLE{Merkys2016,
  author = {Merkys, Andrius and Vaitkus, Antanas and Butkus, Justas and Okulič-Kazarinas,
	Mykolas and Kairys, Visvaldas and Gražulis, Saulius},
  title = {{{\it COD::CIF::Parser}: an error-correcting CIF parser for the Perl
	language}},
  journal = {Journal of Applied Crystallography},
  year = {2016},
  volume = {49},
  number = {1},
  month = {Feb},
  doi = {10.1107/S1600576715022396},
  url = {https://doi.org/10.1107/S1600576715022396}
}

@ARTICLE{Grazulis2015,
  author = {Gražulis, Saulius and Merkys, Andrius and Vaitkus, Antanas and Okulič-Kazarinas,
            Mykolas},
  journal = {Journal of Applied Crystallography},
  pages = {85-91},
  title = {Computing stoichiometric molecular composition from crystal structures},
  year = {2015},
  month = {Feb},
  number = {1},
  volume = {48},
  doi = {10.1107/S1600576714025904},
  url = {https://doi.org/10.1107/S1600576714025904}
}

@article{Grazulis2012,
author = {Gražulis, Saulius and Daškevič, Adriana and Merkys, Andrius and Chateigner, Daniel and Lutterotti, Luca and Quirós, Miguel and Serebryanaya, Nadezhda R. and Moeck, Peter and Downs, Robert T. and Le Bail, Armel}, 
title = {Crystallography Open Database (COD): an open-access collection of crystal structures and platform for world-wide collaboration}, 
volume = {40}, 
number = {D1}, 
pages = {D420-D427}, 
year = {2012}, 
doi = {10.1093/nar/gkr900}, 
URL = {https://academic.oup.com/nar/article/40/D1/D420/2903497}, 
eprint = {https://nar.oxfordjournals.org/content/40/D1/D420.full.pdf+html}, 
journal = {Nucleic Acids Research} 
}

@article{Grazulis2009,
author = "Gra{\v{z}}ulis, Saulius and Chateigner, Daniel and Downs, Robert T. and Yokochi, A. F. T. and Quir{\'{o}}s, Miguel and Lutterotti, Luca and Manakova, Elena and Butkus, Justas and Moeck, Peter and Le Bail, Armel",
title = "{Crystallography Open Database {--} an open-access collection of crystal structures}",
journal = "Journal of Applied Crystallography",
year = "2009",
volume = "42",
number = "4",
pages = "726--729",
month = "Aug",
doi = {10.1107/S0021889809016690},
url = {https://doi.org/10.1107/S0021889809016690},
}

@ARTICLE{Downs2003,
  author = {Downs, R. T. and Hall-Wallace, M.},
  title = {The American Mineralogist Crystal Structure Database},
  journal = {American Mineralogist},
  year = {2003},
  volume = {88},
  pages = {247-250},
}

@article{szymanski2024integrated,
  title={Integrated analysis of X-ray diffraction patterns and pair distribution functions for machine-learned phase identification},
  author={Szymanski, Nathan J and Fu, Sean and Persson, Ellen and Ceder, Gerbrand},
  journal={Npj computational materials},
  volume={10},
  number={1},
  pages={45},
  year={2024},
  publisher={Nature Publishing Group UK London}
}

@article{fei2025dara,
  title={Dara: Automated multiple-hypothesis phase identification and refinement from powder X-ray diffraction},
  author={Fei, Yuxing and McDermott, Matthew J and Rom, Christopher L and Wang, Shilong and Ceder, Gerbrand},
  journal={arXiv preprint arXiv:2510.19667},
  year={2025}
}

@book{cullity2001xrd,
  title        = {Elements of X-Ray Diffraction},
  author       = {Cullity, B. D. and Stock, S. R.},
  edition      = {3rd},
  year         = {2001},
  publisher    = {Prentice Hall},
  address      = {Upper Saddle River, NJ},
  isbn         = {978-0201610918}
}

@article{tamura2022automatic,
  title={Automatic Rietveld refinement by robotic process automation with RIETAN-FP},
  author={Tamura, Ryo and Sumita, Masato and Terayama, Kei and Tsuda, Koji and Izumi, Fujio and Matsushita, Yoshitaka},
  journal={Science and Technology of Advanced Materials: Methods},
  volume={2},
  number={1},
  pages={435--444},
  year={2022},
  publisher={Taylor \& Francis}
}

@misc{segal2025losslandscapepowderxray,
      title={The Loss Landscape of Powder X-Ray Diffraction-Based Structure Optimization Is Too Rough for Gradient Descent}, 
      author={Nofit Segal and Akshay Subramanian and Mingda Li and Benjamin Kurt Miller and Rafael Gomez-Bombarelli},
      year={2025},
      eprint={2512.04036},
      archivePrefix={arXiv},
      primaryClass={cond-mat.mtrl-sci},
      url={https://arxiv.org/abs/2512.04036}, 
}

@article{jette1933x,
  title={An x-ray study of the W{\"u}stite (FeO) solid solutions},
  author={Jette, Eric R and Foote, Frank},
  journal={The Journal of Chemical Physics},
  volume={1},
  number={1},
  pages={29--36},
  year={1933},
  publisher={American Institute of Physics}
}

@book{montoro1938miscibilita,
  title={Miscibilita fra gli ossidi salini di ferro e di manganese},
  author={Montoro, VINCENZO},
  year={1938}
}

\end{document}


\author{Frank M. Abel}
\affiliation{Physics Department, United States Naval Academy, Annapolis, MD 21402, USA}
\affiliation{National Institute of Standards and Technology, Gaithersburg, MD 20899, USA}

\author{Paige Burke}
\affiliation{Physics Department, United States Naval Academy, Annapolis, MD 21402, USA}

\author{Daniel Wines}
\affiliation{National Institute of Standards and Technology, Gaithersburg, MD 20899, USA}

\author{\\Brian Donovan}
\affiliation{Physics Department, United States Naval Academy, Annapolis, MD 21402, USA}

\author{Michelle E. Jamer}
\affiliation{Physics Department, United States Naval Academy, Annapolis, MD 21402, USA}

\author{Kamal Choudhary}
\affiliation{Material Science and Engineering,\\ 
Johns Hopkins University, Baltimore, MD, 21218, USA}
\affiliation{National Institute of Standards and Technology, Gaithersburg, MD 20899, USA}

\begin{center}
\textbf{Supplementary Information for\\ ``Machine Learning for Predicting Magnetization from X-ray Diffraction of Iron Oxide Nanoparticles Using Simple Physics-Based Data Generation"}
\maketitle
\end{center}

\section*{Magnetic Simulation Constants and Model Training Parameters}

\begin{table}[ht]
\centering
\caption{Magnetic simulation constants used in the iron oxide model.}
\label{tab:TabS1}
\renewcommand{\arraystretch}{1.15}
\begin{tabular}{@{}l c@{}}
\toprule
\textbf{Parameter} & \textbf{Value} \\
\midrule
$M_{s,\mathrm{bulk}}$                    & $92\,\mathrm{A\,m^2\,kg^{-1}}$ \\
$\chi_{\mathrm{FeO}}$ (mass susceptibility) & $1\times 10^{-6}\,\mathrm{m^3\,kg^{-1}}$ \\
$\rho_{\mathrm{Fe_3O_4}}$                & $5.2\times 10^{3}\,\mathrm{kg\,m^{-3}}$ \\ 
$\mu_0$                                   & $1.25663706127\times 10^{-6}\,\mathrm{N\,A^{-2}}$ \\
$k_B$                                     & $1.380649\times 10^{-23}\,\mathrm{J\,K^{-1}}$ \\
$T$                                       & $300\,\mathrm{K}$ \\
$t$ (surface layer thickness)             & $0.3\,\mathrm{nm}$ \\
$V$ (particle volume)                     & $\displaystyle \frac{4}{3}\pi\!\left(\frac{D}{2}\right)^{3}$ \\
$B_{\min}$ (magnetic field)               & $-2.8\,\mathrm{T}$ \\
$B_{\max}$ (magnetic field)               & $+2.8\,\mathrm{T}$ \\
\bottomrule
\end{tabular}
\end{table}

\begin{table}[ht]
\begin{center}    
\caption{Model hyperparameters used for Random Forest (RF) and Gradient Boosting (GB), which obtained $>$ 0.9 R$^2$ metric and are the primary models examined and validated in the main text. For the ``max\_features'' ``sqrt", this means the square root is taken of the number of features, which is 2000 for both models (number of intensity values).}
\label{tab:TabS2}
\renewcommand{\arraystretch}{1.15}
\begin{tabular}{@{}lcc@{}}
\toprule
\textbf{Hyperparameter} & \textbf{RF} & \textbf{GB} \\
\midrule
n\_estimators        & $500$        & $800$ \\
learning\_rate       & --                  & $0.03$ \\
max\_depth           & --                  & $3$ \\
min\_samples\_leaf   & $3$          & $5$ \\
max\_features        & ``sqrt''     & ``sqrt'' \\
subsample            & --                  & $0.8$ \\
loss                 & --                  & ``huber'' \\
alpha                & --                  & $0.9$ \\
validation\_fraction & --                  & $0.1$ \\
n\_iter\_no\_change  & --                  & $20$ \\
tol                  & --                  & $1 \times 10^{-4}$ \\
random\_state        & $1$          & $1$ \\
n\_jobs              & $-1$         & -- \\
\bottomrule
\end{tabular}
\end{center}
\end{table}

\clearpage

\section*{Relation Between Magnetic Properties and Shared Parameters for Simulated Datasets}

\begin{figure*}[h]
    \centering
    \includegraphics[width = 1.0\textwidth]{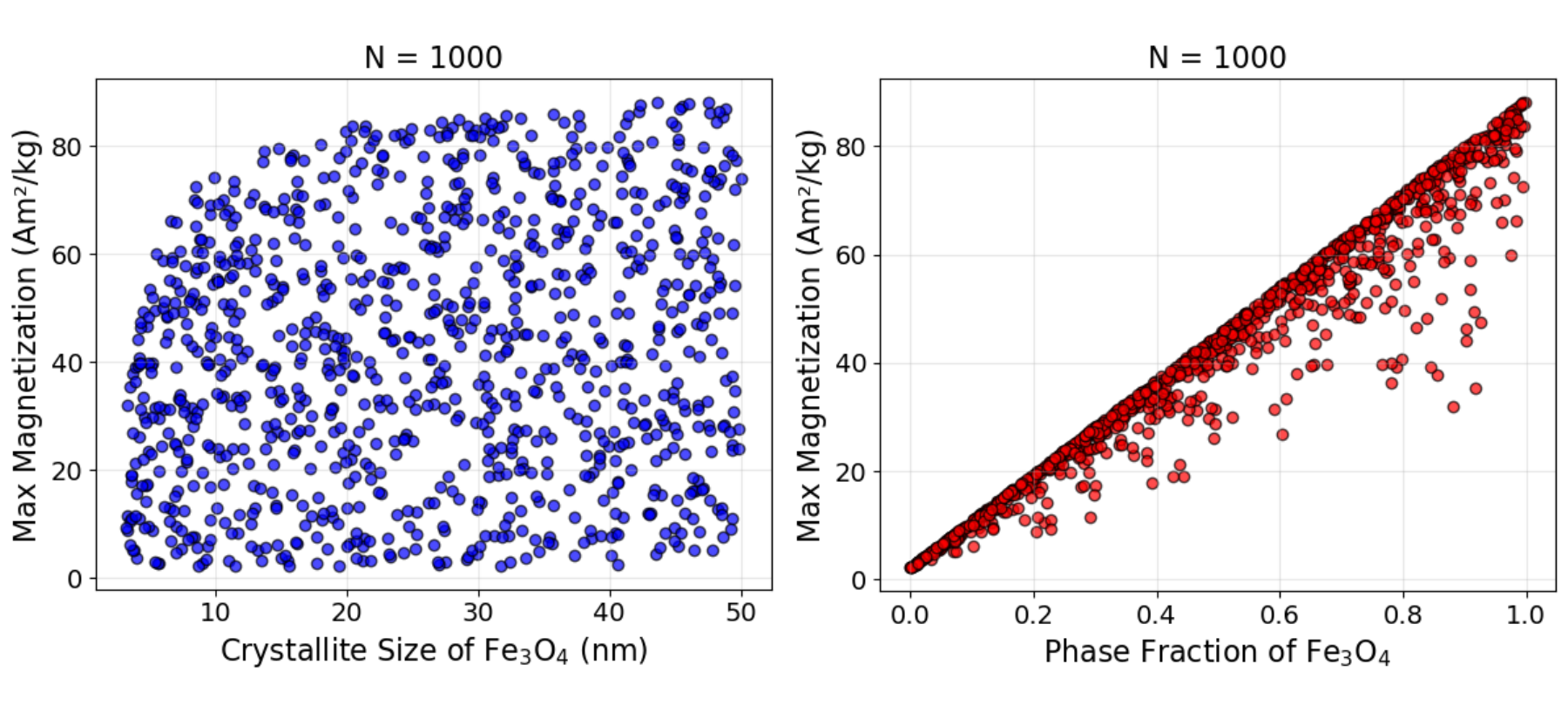}
     \caption{Max magnetization (at 2.8 T) as a function of crystallite size and phase fraction Fe$_3$O$_4$ for a dataset of size N = 1000 using a random seed of ``42''.}
    \label{fig:FigS1}
\end{figure*}

\clearpage

\section*{Effect of Dataset Size, Hyperparameters, and Random Seed on Model Predictions}

\begin{figure*}[h]
    \centering
    \includegraphics[width = 1.0\textwidth]{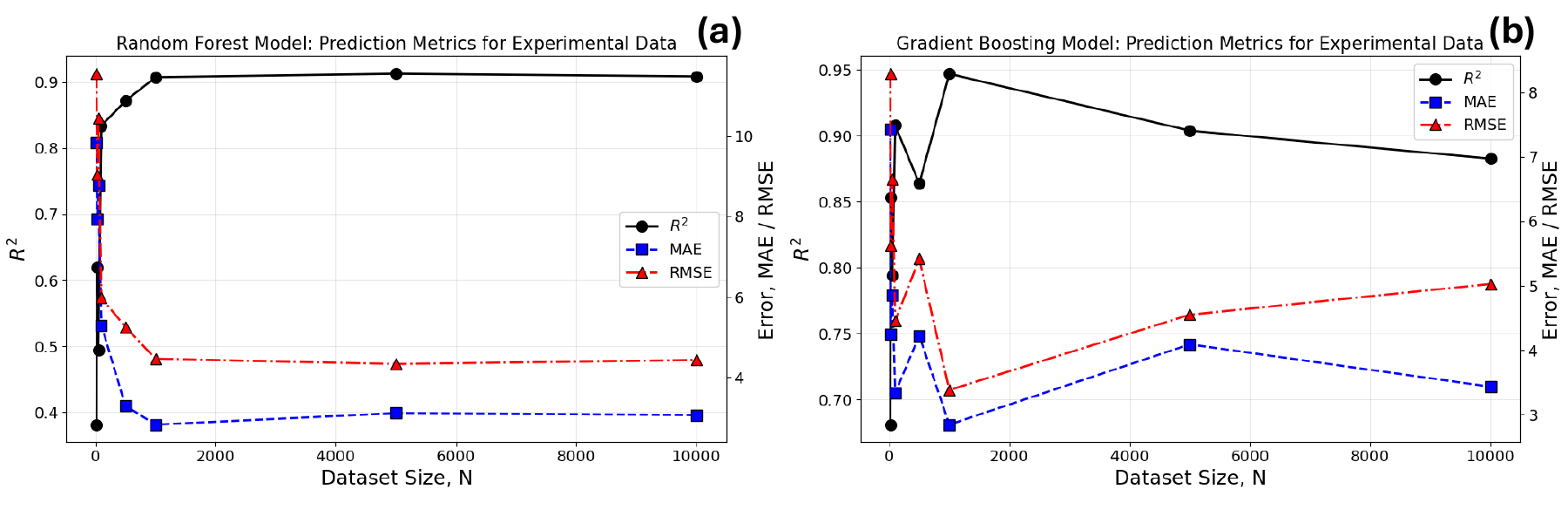}
     \caption{Effect of simulated dataset size on prediction metrics for experimental samples, RF model (a) and GB model (b). Models correspond to hyperparameter given in Table S2.}
    \label{fig:FigS2}
\end{figure*}

\begin{figure*}[h]
    \centering
    \includegraphics[width = 1.0\textwidth]{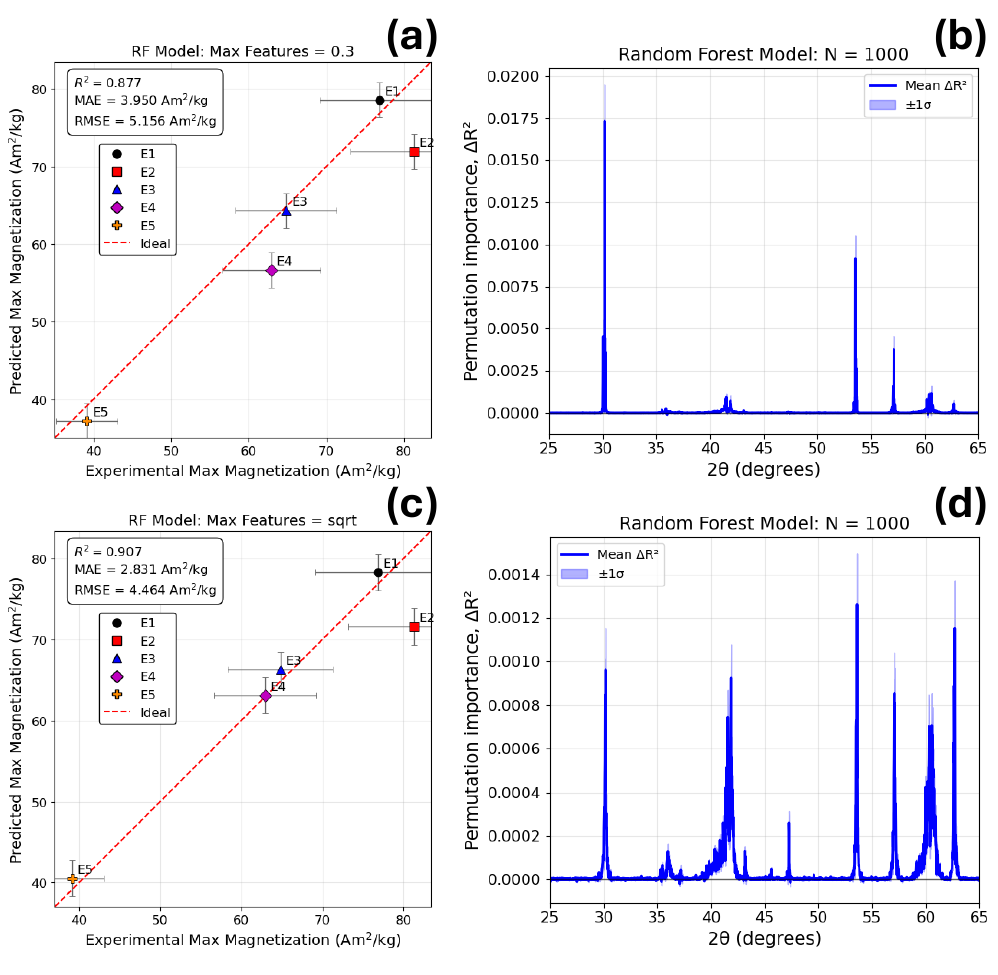}
     \caption{Comparing the change in the ``max\_feature'' hyperparameter of the prediction of max magnetization on the experimental samples and corresponding feature importance for the RF model; 0.3 corresponding to 600 max features (a, b), and ``sqrt'' corresponding to about 44.7 max features (c, d).}
         \label{fig:FigS3}
\end{figure*}

\begin{figure*}[h]
    \centering
    \includegraphics[width = 1.0\textwidth]{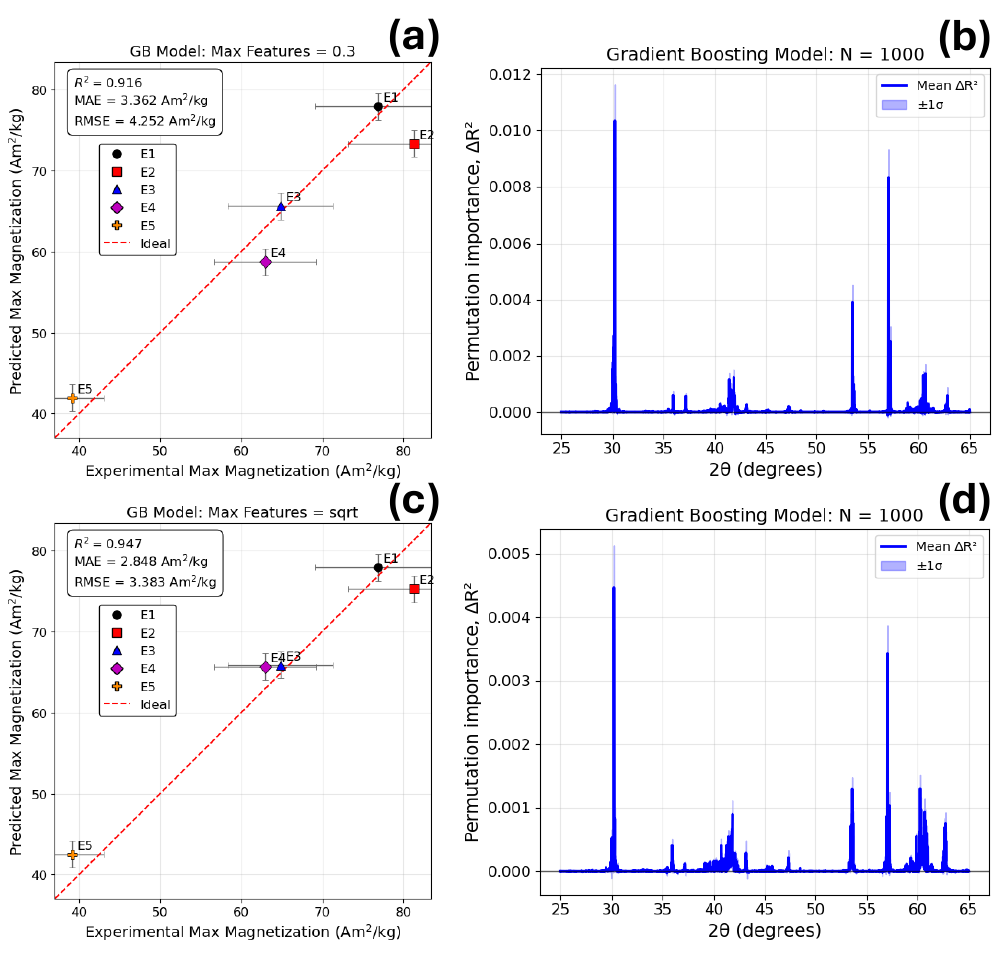}
     \caption{Comparing the change in the ``max\_feature'' hyperparameter of the prediction of max magnetization on the experimental samples and corresponding feature importance for the GB model; 0.3 corresponding to 600 max features (a, b), and ``sqrt'' corresponding to about 44.7 max features (c, d).}
         \label{fig:FigS3}
\end{figure*}

\begin{table}[ht]
\centering
\caption{Summary of RF and GB model prediction metrics and the effect of the ``max\_features'' hyperparameter trained on simulated datasets (N = 1000) with different random seeds, and mean and standard deviation (SD) of the four seeds. With 0.3 corresponding to 600 max features, and ``sqrt'' corresponding to about 44.7 max features, the features being the array of 2000 intensity values for a given simulated XRD pattern. The units of mean average error (MAE), root mean square error (RMSE) are in (A\,m$^2$\,kg$^{-1}$).}
\label{tab:TabS3}

\scriptsize
\renewcommand{\arraystretch}{1.0}

\begin{tabular}{@{}l c c c c c@{}}
\toprule
\textbf{Model} & \textbf{Seed} & \textbf{max\_features} & \textbf{R$^2$} & \textbf{MAE} & \textbf{RMSE}  \\
\midrule
RF & 1  & 0.3     & 0.631 & 7.389 & 8.920 \\
RF & 1  & ``sqrt" & 0.853 & 3.373 & 5.635 \\
RF & 22 & 0.3     & 0.857 & 5.106 & 5.551 \\
RF & 22 & ``sqrt" & 0.900 & 2.484 & 4.650 \\
RF & 42 & 0.3     & 0.877 & 3.950 & 5.156 \\
RF & 42 & ``sqrt" & 0.907 & 2.831 & 4.464 \\
RF & 80 & 0.3     & 0.770 & 6.026 & 7.042 \\
RF & 80 & ``sqrt" & 0.906 & 2.701 & 4.496 \\
\midrule
GB & 1  & 0.3     & 0.859 & 4.247 & 5.513 \\
GB & 1  & ``sqrt" & 0.906 & 2.897 & 4.492 \\
GB & 22 & 0.3     & 0.928 & 3.022 & 3.950 \\
GB & 22 & ``sqrt" & 0.934 & 3.140 & 3.774 \\
GB & 42 & 0.3     & 0.916 & 3.362 & 4.252 \\
GB & 42 & ``sqrt" & 0.947 & 2.848 & 3.383 \\
GB & 80 & 0.3     & 0.802 & 4.963 & 6.533 \\
GB & 80 & ``sqrt" & 0.931 & 3.214 & 3.841 \\
\bottomrule
\end{tabular}

\vspace{0.75em}

\begin{tabular}{@{}l c c@{\hspace{0.8em}}c c@{\hspace{0.8em}}c c@{\hspace{0.8em}}c@{}}
\toprule
\textbf{Model} & \textbf{max\_features} &
\multicolumn{2}{c}{$R^2$} &
\multicolumn{2}{c}{MAE} &
\multicolumn{2}{c}{RMSE} \\
\cmidrule(lr){3-4} \cmidrule(lr){5-6} \cmidrule(lr){7-8}
 & & \textbf{Mean} & \textbf{SD} & \textbf{Mean} & \textbf{SD} & \textbf{Mean} & \textbf{SD} \\
\midrule
RF & 0.3           & 0.783 & 0.112  & 5.618 & 1.455 & 6.667 & 1.707 \\
RF & ``sqrt"       & 0.891 & 0.026  & 2.847 & 0.379 & 4.811 & 0.555 \\
GB & 0.3           & 0.876 & 0.058  & 3.899 & 0.8776 & 5.062 & 1.192 \\
GB & ``sqrt"       & 0.930 & 0.017  & 3.025 & 0.1796 & 3.873 & 0.4598 \\
\bottomrule
\end{tabular}

\vspace{0.75em}

\vspace{0.75em}
\begin{tabular}{@{}l c@{\hspace{0.3em}}c@{}}
\toprule
\textbf{Model} & \textbf{Metric} & \textbf{Mean $\Delta$} \\
\midrule
RF & $R^2$  & 0.108 \\
GB & $R^2$  & 0.054 \\
RF & MAE    & 2.771 \\
GB & MAE    & 0.8735 \\
RF & RMSE   & 1.856 \\
GB & RMSE   & 1.189 \\
\midrule
\multicolumn{3}{@{}l@{}}{\footnotesize
$\Delta$ denotes the change when using max\_features = ``sqrt'' instead of 0.3.} \\
\multicolumn{3}{@{}l@{}}{\footnotesize
$\Delta\Delta R^2$ (RF improvement $-$ GB improvement) $\approx 0.054$;} \\
\multicolumn{3}{@{}l@{}}{\footnotesize
$\Delta\Delta \mathrm{MAE} \approx 1.9$ A\,m$^2$\,kg$^{-1}$, \quad
$\Delta\Delta \mathrm{RMSE} \approx 0.67$ A\,m$^2$\,kg$^{-1}$.} \\
\bottomrule
\end{tabular}

\end{table}

\section*{Effect of XRD Measurement Time}

\begin{figure*}[h]
    \centering
    \includegraphics[width = 1.0\textwidth]{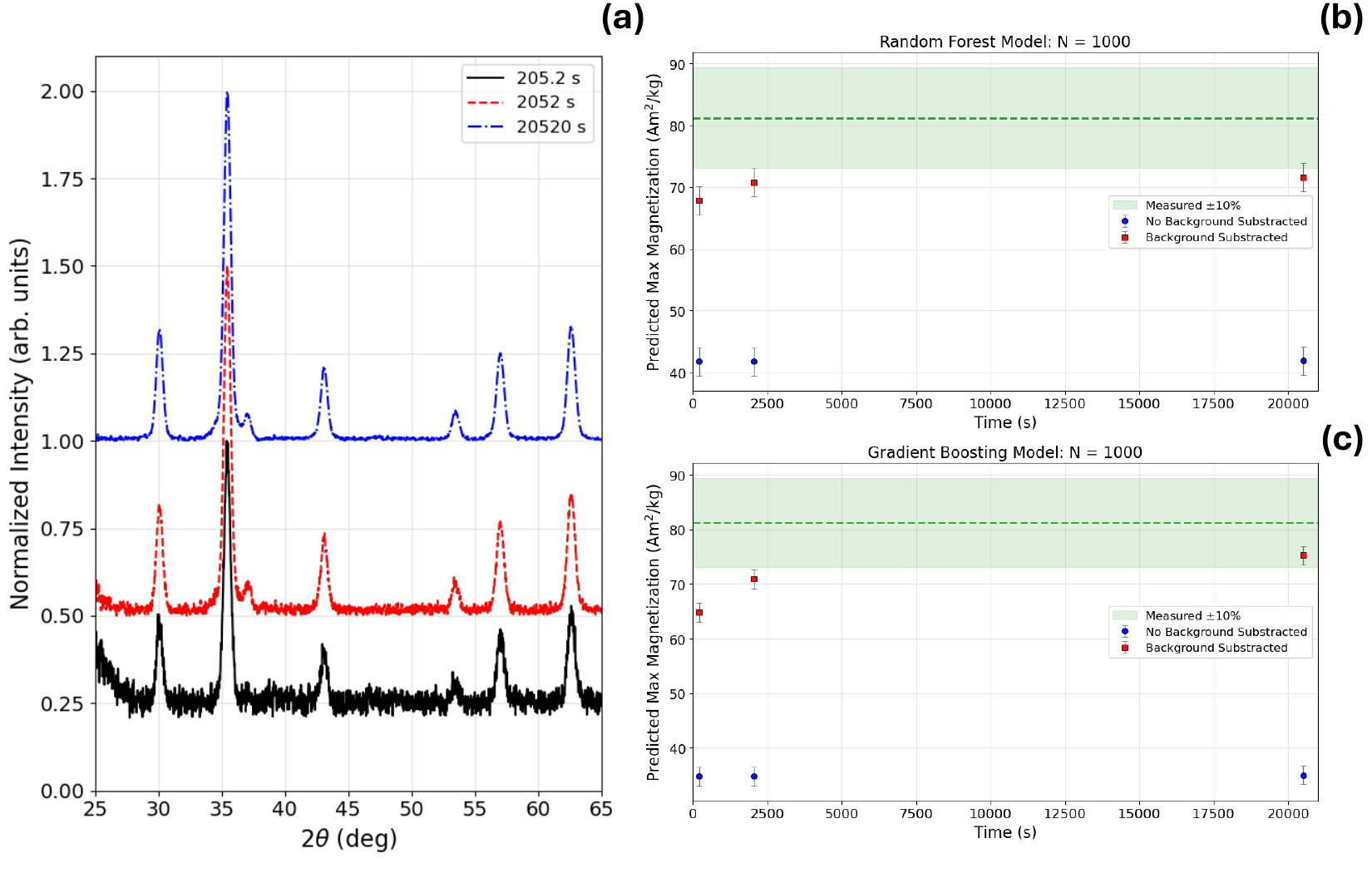}
     \caption{XRD of sample S2 with three different measurement times 205.2 s, 2052 s, and 20520 s (a), RF model predictions for the three different measurement times, with and without background subtraction (b), and GB model predictions for the three different measurement times, with and without background subtraction (c).  Models correspond to the hyperparameters given in Table S2.}
         \label{fig:FigS5}
\end{figure*}

\section*{Iron Oxide Nanoparticle Synthesis}

The iron oxide nanoparticles were all synthesized by thermal decomposition approaches of Fe(acac)$_3$ in a mixture of organic solvents. \cite{sun2002size, castellanos2021milestone} Four of the samples which had large majorities of Fe$_3$O$_4$/$\gamma$-Fe$_2$O$_3$ were synthesized by first creating a stock complex of Fe(acac)$_3$, 1,2-hexadecanediol, 1-Octadecene, and Oleic acid. To perform a given synthesis, the stock complex is heated to about 50 $^o$C, and the stock complex is then added along with Benzyl ether and 1-Octadecene by micropipette into a 3-stopper flask, after which a heating profile is performed. One of these samples had Zn(acac)$_2$ added to the reaction flask before the addition of the three solutions by micropipette. The sample with a mixture of FeO and Fe$_3$O$_4$/$\gamma$-Fe$_2$O$_3$ was produced using a 50/50 mixture of Oleic acid and Oleylamine with Fe(acac)$_3$ in a 3-stopper flask, followed by a heating profile. The sample initially had mostly FeO \cite{khurshid2013synthesis}, but was aged for about five and a half months, resulting in a mixture of FeO and Fe$_3$O$_4$/$\gamma$-Fe$_2$O$_3$, after which the XRD and magnetization vs. magnetic field data were taken that are reported in this work.

\clearpage
\label{References}
\bibliographystyle{apsrev4-1}
\bibliography{bib}